\documentclass[twocolumn,showpacs,preprintnumbers,amsmath,amssymb,prb]{revtex4}
\usepackage[utf8]{inputenc}
\usepackage{amsmath}
\usepackage{subfigure}
\usepackage{array}
\usepackage{graphicx}
\usepackage{fancyheadings}
\usepackage{dcolumn}
\usepackage{bm}
\usepackage{graphicx,color}

\def\be{\begin{equation}}
\def\ee{\end{equation}}
\def\ba{\begin{eqnarray}}
\def\ea{\end{eqnarray}}

\begin{document}

\title{Equation of state of charged colloidal suspensions and its 
dependence on the thermodynamic route}

\author{Thiago E. Colla}

\author{Alexandre P. dos Santos}

\author{Yan Levin}
\affiliation{
Instituto de F\'isica, Universidade Fedaral do Rio Grande do Sul, CP 15051,
91501-970 Porto Alegre, RS, Brazil.
}

\begin{abstract}
The thermodynamic properties of highly charged colloidal suspensions in
contact with a salt reservoir are investigated in the framework of the
Renormalized Jellium Model (RJM). It is found that the equation of state 
is very sensitive to the particular thermodynamic route used to obtain it. 
Specifically, the osmotic pressure calculated within the RJM using the
contact value theorem can be very 
different from the pressure calculated  
using the Kirkwood-Buff fluctuation relations. On the other hand,
Monte Carlo (MC) simulations show that both the effective 
pair potentials
and the correlation functions are accurately predicted by the RJM. It is
suggested
that the lack of self-consistency in the thermodynamics of the RJM is a result 
of neglected electrostatic correlations between the counterions and coions.

\end{abstract}

\maketitle

\section{Introduction}

In spite of the fundamental importance --- both practical and theoretical ---
the thermodynamic properties of charged colloidal suspensions are far
from understood \cite{Rev_Lev,Rev_Mes,Rev_Bell,Mah_00}. Even such basic question
as the existence of a liquid-gas phase transition in 
these systems still remains a topic of debate
\cite{Die_01,Lev_03,Den_07,Ben_07}. The
difficulty in describing the thermodynamics of charged colloidal suspensions
is a consequence of both
size and charge asymmetry between the different components of the system
and the long-range nature of the
Coulomb interaction \cite{Rev_Lev,Rev_Han,Ver_11}. To simplify the theoretical
description  one often uses the, so-called, Primitive Model (PM).  In this model
 all
charged components --- colloidal particles, coions, and counterions --- are
treated explicitly, while the solvent --- usually an aqueous medium --- is
considered 
as a dielectric continuum. The interactions between the colloidal particles,
the counterions, and the coions have both Coulomb and hard-core components.
Image effects resulting from the dielectric discontinuities across the particle
surface are usually neglected at the lowest order of approximation. 

Colloidal suspensions often contain salt.  For theoretical description it 
is, therefore, convenient to work in a semi-grand-canonical ensemble in which 
the  number of colloidal particles is fixed, while the concentration of
salt is controlled by an externally imposed chemical potential.
Physically this can be realized by separating the suspension from a salt
reservoir by a 
semi-permeable membrane transparent only to microions
\cite{Tam_98,Des_02}.

The large asymmetry between the colloidal particles and the
microions, forces us to employ different approximations to account for
the correlations among the various components of suspension. 
The correlations among the microions
can be described by a linear Debye-H\"uckel (DH) like theory.  
For dilute colloidal suspensions these correlations are 
usually negligible.  On the other
hand, to account for strong colloid-ion and colloid-colloid
interactions requires a full non-linear theory. One approach that
has proven to be very useful for describing the non-linear correlations
between the colloidal particles and the counterions is the
concept of charge renormalization \cite{Rev_Lev,Bell_98,Gon_08}. 
The idea is that
strong electrostatic attraction between the colloidal particles and their
counterions will lead to 
accumulation
of counterions near the colloidal surface.  These counterions can be
considered to be "condensed" (strongly bound) to the colloidal particle,
effectively
renormalizing its bare charge. 
For strongly charged colloidal particles the
renormalized charge will, in general, be much smaller in magnitude than the bare
charge \cite{Bell_98}. 

An alternative, but equivalent way of modeling colloidal suspensions is
to
explicitly trace out the microion degrees of freedom in a semi-grand-canonical
partition function \cite{Rev_Lik}. This way the multi-component colloidal
suspension
is mapped onto an equivalent one-component
system in which only the colloidal particles are explicit. 
This coarse-graining procedure defines the, so-called, One Component
Model (OCM). In
this approach, all the contributions coming from the traced-out microions are 
implicit in the effective interactions between the
colloidal particles \cite{Rev_Lik}. The apparent simplification over the
original problem is only formal, since the effective interaction 
between the colloidal particles now has a many-body character
\cite{Rev_Lik,Rev_Den,Den_03}
and is state-dependent \cite{Rev_Lik,Rev_Den},  further
complicating the thermodynamic calculations \cite{Sti_02,Louis_02,Dijk_00,Emm_07}. 

For weakly charged colloidal particles, the effective interaction potential 
in the OCM takes
a particularly simple form known as the
Derjaguin-Landau-Overbeek-Verwey (DLVO) pair potential \cite{DLVO1,DLVO2}, 
\begin{equation}
\label{eq1}
\beta u(r)=\lambda_{B}\left(\dfrac{Z e^{\kappa a}}{1+\kappa a
}\right)^{2}\dfrac{e^{-\kappa r}}{r},
\end{equation}
where $a$ and $-Z q$ are the colloidal radius and charge, respectively.
The inverse Debye screening length is 
$\kappa^{2}\equiv \sqrt{4\pi\lambda_{B} (\rho_{+}+\rho_{-})}$, where
$ \rho_{+}$ and $\rho_{-}$ are the mean 
concentrations of the monovalent counterions and coions inside the suspension,
and 
$\lambda_{B}\equiv\beta q^{2}/\epsilon$ is the Bjerrum length.  Due to the
global charge neutrality, $\rho_{+}-\rho_{-}-Z\rho=0$, where $\rho$ is the
concentration of colloidal particles. 

For strongly charged colloidal particles, the linear DLVO theory is not
sufficient
to describe the pairwise interactions.  The non-linear effects, however, can
be included into DLVO potential by explicitly accounting for the counterion
condensation. This can be achieved by
replacing the bare colloidal charge in Eq.~(\ref{eq1}) by the
renormalized effective
charge $Z\rightarrow Z_{eff}$. The charge renormalization 
accounts for the 
strong non-linear particle-counterion correlations near
the colloidal surfaces. 

Besides the DLVO pair potential, the effective colloidal interactions in the OCM
formalism also have the, so-called, volume terms which 
depend on colloidal density, but not on colloidal coordinates
\cite{Rev_Bell,Rev_Han}.  The volume terms
were argued to play important role for the thermodynamics of charged colloidal
suspensions
\cite{Rev_Lik,Bas&Roij_06,Chan_01}.
For structural properties of the OCM, however, these terms
do not play any role, since they do not depend on colloidal coordinates. This
point must be considered with special care when one wants to study colloidal
thermodynamics using the OCM \cite{Rev_Lik,Dijk_00,Emm_07}. In fact,
there are some approaches that describe the effective interaction by simply
\textit{defining}
a pair
potential which reproduces the correct colloid-colloid correlations in
using the OCM. Clearly, such approaches must loose some 
thermodynamic informations contained in
the volume terms.

The question of whether the 
effective potential models based purely on  pair interactions are
sufficient to study the thermodynamics of a
fully multi-component system is still under discussion
\cite{Louis_02,Sti_02}. In the case of
charged colloidal systems, the problem is even more subtle, since such
systems must obey additional constraints {\it i.e.}  global electro-neutrality
and the
well-known Stillinger-Lovett moment conditions \cite{SL}. As a consequence, many
theoretical tools originally designed for unconstrained systems have to be
reformulated before they can be applied to charged systems
\cite{Fri_70,Kus_87}.

The aim of this work is to address some thermodynamic inconsistencies which
arise when different
routes are used to calculate the thermodynamic functions of charged
colloidal suspensions. To this end, we will use the  Renormalized
Jellium Model (RJM), from which both the renormalized charge and the osmotic pressure
can be easily calculated \cite{Emm_04,Sal_07}. From the renormalized charge,
the effective pair
potential --- and hence the colloid-colloid pair correlation functions --- can be
obtained
using the OCM Ornstein-Zernike (OZ) equation
\cite{Hansen} with an appropriate closure.
Knowing the correlations, it is possible to calculate the osmotic
compressibility
using the Kirkwood-Buff (KB) fluctuation theory \cite{KB}. 
In this work we will compare the osmotic compressibilities of the RJM calculated
using both the contact theorem and the KB fluctuation relations.

The paper is organized as follows. In section II we will briefly review 
the theoretical methods used
for the thermodynamic investigations ---  the RJM, the
Donnan Equilibrium, and the Kirkwood-Buff relations. In
section III, we will briefly 
discuss the simulation techniques employed in this study. 
The results will be presented in section IV, 
and conclusions, discussion, and suggestions for the
future investigations will be given in section V.

                      \section{Theoretical Background}

\subsection{The Renormalized Jellium Model}

The RJM is a model that allows one to calculate the effective charge of
colloidal
particles and the
thermodynamic properties of colloidal suspensions based on
the mean-field
Poisson-Boltzmann-like (PB) equation. RJM is known to be very 
accurate for salt-free 
colloidal suspensions with monovalent counterions. In contrast to
the traditional Cell Model (CM), where a lattice-like structure is assumed for
colloidal particles, in the RJM the colloidal correlations are modeled by
a uniform neutralizing background. The major conceptual advantage of the RJM
over the CM is that the pair potential Eq.~(\ref{eq1}) is exact within the
RJM formalism, while for CM there is no pairwise interaction between the
colloidal particles \cite{Emm_04}.  Thus, 
the effective charges calculated using CM have no clear
connection with the DLVO potential.
Recently, the RJM was successfully extended to incorporate 
inter-colloidal correlations \cite{Cas_06,Col_09}, the multivalent
counterions
\cite{Col_10}, and colloidal  
polydispersity \cite{Fal_11}.

In the RJM, one colloidal particle  of charge $-Z_{bare}q$ and radius $a$ 
is fixed at the
origin of the coordinate system.  The distribution of
\textit{free} (uncondensed) ions around this particle is assumed to 
follow the Boltzmann
distribution, $\rho_{\pm}(r)=\rho_{\pm}e^{\mp \beta q \psi(r)}$, where 
$\rho_{\pm}$ are the counterion and coion mean densities, and $\psi(r)$ is
the mean electrostatic potential. The remaining colloidal particles, 
along with their condensed counterions, are taken to provide a 
uniform neutralizing background of charge density $-Z_{eff} q \rho$. The
reduced mean electrostatic potential $\phi(r)=\beta q \psi(r)$ then satisfies
the
Poisson-Boltzmann-Jellium (PBJ) equation:
\begin{equation}
\label{eq2}
\nabla^{2}
\phi(r)=-4\pi\lambda_{B}\left(\rho_{+}e^{-\phi(r)}-\rho_{-}e^{\phi(r)}-Z_{eff}
\rho\right).
\end{equation}
This equation can be numerically solved with the boundary conditions
$\phi(r\rightarrow \infty)\rightarrow 0$ and
$\frac{d\phi(r)}{dr}\arrowvert_{r=a}=\frac{Z_{bare}\lambda_{B}}{a^{2}}$. The
first condition defines the zero of the electrostatic potential in the bulk of suspension,
while the second one determines the electric field at the colloidal surface using the Gauss law. 

Far from the
central colloidal particle  --- the region where the electrostatic potential is
weak ---  the PBJ equation can be linearized, resulting in the
following long-distance behavior:
\begin{equation}
\label{eq3}
\phi(r)=-\dfrac{Z_{eff}\lambda_{B}e^{\kappa a}}{(1+\kappa a)}\dfrac{e^{-\kappa r
}}{r},
\end{equation}
where
$\kappa=\sqrt{4\pi\lambda_{B}(\rho_{+}+\rho_{-})}=\sqrt{4\pi\lambda_{B}(2\rho_{-
}+Z_{eff}}
\rho)$ is the \textit{effective} screening length, and where we have used the
global charge
neutrality condition $\rho_{+}-\rho_{-}-\rho Z_{eff}=0$. Note that in the
far-field, the bare charge $Z_{bare}$ is replaced by the renormalized
charge $Z_{eff}$, reflecting the nonlinear correlations at the colloidal surface. 

For a given salt and colloidal concentrations, $\rho_{-}$ and $\rho$, respectively,
the effective charge is calculated by matching the  numerical solution of
Eq.~(\ref{eq2}) with the linearized potential Eq.~(\ref{eq3}), in the far-field.
Since within the RJM the background charge arises from the smeared-out charge of
colloidal particles and their condensed counterions, the self-consistency
requires that the effective colloidal charge must be the same as 
the charge of the uniform neutralizing background.   This procedure can 
be easily implemented numerically \cite{Fal_10}. Suppose that we know $Z_{eff}$,
then from  Eq. (\ref{eq3}) we will also know the potential and the electric
field
in the far-field region.  We can then integrate the PBJ equation using a
standard Rounge-Kutta algorithm to obtain the electrostatic potential all the
way up to the colloidal
surface.   The
corresponding bare colloidal
charge $Z_{bare}$ is obtained using the Gauss low at the
colloidal surface. In reality, of course, one wants to 
calculate the effective charge for
a given bare charge.  This can be done by varying $Z_{eff}$ until
the desired $Z_{bare}$ is found.  In practice, this can be easily implemented
numerically
by incorporating a Newton-Raphson root-finding subroutine in the PBJ solver.

The osmotic pressure within the RJM is given by
\begin{equation}
\label{eq4}
\beta P = \rho_{+}+\rho_{-},
\end{equation}
where $\rho_{\pm}$ are the bulk concentrations of \textit{free} coions and
counterions. In spite of its apparent simplicity, this ideal-gas-like equation
of
state requires a knowledge of  microion concentrations in the far-field which, in 
turn, depend on the charge
renormalization and osmotic equilibrium with the salt reservoir.  We should
also note that unlike for CM, for which the contact value theorem is 
an exact statement \cite{Wen_82,Des_01,Car_81,Mar_55,Des_01}, Eq. (\ref{eq4}) of
the RJM is only valid in the mean-field approximation.
We will later argue that the failure to properly account for ionic correlations
leads to thermodynamics inconsistencies in the RJM.

\subsection{The Donnan Equilibrium}

In this work we will consider a colloidal suspension in contact with a salt
reservoir.  The system is separated from the reservoir by 
a semi-permeable membrane which allows for a free flux of  microions. 
The ionic concentration inside the suspension will then be determined
by the osmotic equilibrium with the salt reservoir. Contrary to uncharged
systems,
for which the osmotic equilibrium simply results in a solvent flow 
 from a solute poor to a solute reach region, 
the osmotic equilibrium in charged
systems is also constrained by the overall charge neutrality of the system.
Physically, this is reflected in the appearance of a
potential difference across the semi-permeable membrane which controls the
overall build
up of charge in the system \cite{Des_02,Tam_98}. This potential
difference is known
as the \textit{Donnan potential} \cite{Donnan}. From a theoretical point of
view, it can also
be thought of as a Lagrange multiplier used to enforce the 
charge neutrality of the system \cite{Tam_98, Bas&Roij_06}.

In equilibrium, the ionic electrochemical potentials inside the system must be
equal to the ones in the salt reservoir. 
Neglecting the electrostatic correlations between the microions, the ionic
concentrations
in the bulk and reservoir are related by $\rho_{\pm}=\rho_{s}e^{\mp \phi_{D}}$,
where $\rho_{s}$ is the salt concentration in the reservoir, and
$\phi_{D}$ is the adimensional Donnan potential. Using the charge neutrality
condition for free ions, $\rho_{+}-\rho_{-}-\rho Z_{eff}=0$, the Donnan
potential
can be eliminated to yield the bulk concentrations of free (uncondensed)
microions:
\begin{equation}
\label{eq5}
\rho_{\pm}=\dfrac{1}{2}\left(\sqrt{(\rho Z_{eff})^{2}+(2\rho_{s})^{2}}\pm
Z_{eff}\rho\right).
\end{equation}

This expression can be used, together with the equation of state
Eq.~(\ref{eq4}), to
write the osmotic pressure $\beta \Pi$ as:
\begin{equation}
\label{JEOS}
\beta \Pi \equiv\beta P-2\rho_{s}=\rho+\sqrt{(\rho
Z_{eff})^{2}+(2\rho_{s})^{2}}-2\rho_{s}.
\end{equation}
where we have also added the colloidal ideal gas contribution $\beta P_c= \rho$.
It is
important to stress that the above expression for the osmotic pressure
completely ignore the microion correlations. This can be justified as long as
the concentration of coions in the bulk is very low. 
The colloid-counterion
correlations are taken into account through the charge renormalization.

Using Eq.~(\ref{JEOS}), two important limits can be verified. For high salt
concentrations  ---  $Z_{eff}\rho/2\rho_{s}\ll1$,  salt-dominated
regime --- there is no significant variation in the ionic concentrations across
the membrane and the osmotic pressure (\ref{JEOS}) is small.  
On the other hand, in the 
limit  $Z_{eff}\rho/2\rho_{s}\gg1$ --- the counterion-dominated
regime --- there is a significant variation in 
the microion concentration between the
bulk suspension and
the reservoir and the osmotic pressure is large \cite{Tam_98}.

The inverse osmotic compressibility $\chi_{T}^{-1}=\rho\left(\dfrac{\partial 
\Pi}{\partial
\rho}\right)_{\rho_{s},T}$ follows directly from
Eq.~(\ref{JEOS}):
\begin{equation}
\label{eq6}
\beta\chi_{T}^{-1}=\rho+\dfrac{\rho^{2} Z_{eff}^{2}}{\sqrt{(\rho
Z_{eff})^{2}+(2\rho_{s})^{2}}}\left(1+\dfrac{d\log(Z_{eff})}{d\log(\eta)}\right)
,
\end{equation}
where $\eta=4\pi a^{3}\rho/3$ is the colloidal volume fraction. The
derivative on the right-hand-side of this expression can be neglected, since in
the RJM the effective charge depends only weakly on the colloidal volume
fraction \cite{Emm_04,Sal_07}.

\subsection{The Kirkwood-Buff relation}

Once the nonlinear colloid-ion correlations are properly taken into account
through the charge renormalization, the DLVO pair potential Eq. (1) can be used
to investigate  the structural properties of the suspension. This can be done
by solving the OCM Ornstein-Zernike equation:
\begin{equation}
\label{OZ}
h(\mathbf{r})=c(\mathbf{r})+\rho\int{h(\mathbf{r}')c(\arrowvert\mathbf{r}
-\mathbf{r}'\arrowvert)d\mathbf{r}'},
\end{equation}
where $h(\mathbf{r})$ and $c(\mathbf{r})$ are the total and the direct correlation
functions, respectively. This equation has to be supplemented by an appropriate
closure relation 
between  $h(\mathbf{r})$ and $c(\mathbf{r})$
\cite{Hansen}.

Once the structural properties are known, the thermodynamic informations can then be
obtained using the Kirkwood-Buff (KB) fluctuation theory
\cite{KB}.
KB theory allows us to express the thermodynamic functions, such as the osmotic
coefficients and the compressibilities, as integrals over the pair
correlation functions. 
Originally formulated for unconstrained mixtures, KB theory requires some
extra care when extended to systems in which the number densities of different
components are not independent \cite{Sti_02,Fri_70,Kus_87}. This is
precisely the case for the
charged systems, for which long-range Coulomb interaction requires 
an overall charge neutrality. 
In addition to this, there are also other constraints 
known as the Stillinger-Lovett
moment conditions, that restrict the fluctuations of different components
\cite{SL} of a charged system. A
naive application of the original KB theory to charged systems leads to
undetermined results \cite{Hansen,Beh_97}. One way of avoiding these
difficulties is to study
the KB integrals for arbitrary $k$ vectors in the Fourier space \cite{Kus_87},
taking the limit $k\rightarrow0$ at the end of the calculations. 
The extended KB theory then relates the osmotic compressibility with the
Fourier transform of the total correlation function $\hat{h}(\mathbf{k})$,
\begin{equation}
\label{KB1}
\chi_{T}=1+\rho\int{h(\mathbf{r})d\mathbf{r}}=1+\rho \hat{h}(0),
\end{equation}
Using OZ equation, this expression can be inverted  to yield
\begin{equation}
\label{KB2}
\left(\dfrac{\partial \beta P}{\partial
\rho}\right)_{\rho_{s},T}=1-\rho\hat{c}(0).
\end{equation}

KB theory shows that the knowledge of colloidal pair correlation function 
is sufficient for calculating the equation of state of the colloidal suspension. 
Curiously, Eqs.~(\ref{KB1})
and (\ref{KB2})
rely only on the pair correlations which are well
modeled using only the effective pair
potential, Eq.~(\ref{eq1}). This suggests that 
the zero-order volume terms, which depend on colloidal concentration
\cite{Rev_Han,Rev_Bell}, are not very important for the thermodynamics. 	

                         \section{Monte Carlo Simulations}

To explore the validity of the RJM model, we perform Monte Carlo simulations
to obtain the "exact" pairwise interaction potential.
The simulations are performed for several fixed distances $R$ between two
spherical
colloidal particles of charge $-Z_{bare}q$, 
which are restricted to move along the main diagonal of a
box of
side length $L=180$\AA.  Colloid particle $1$ is located 
at $x,y,z=-R/2\sqrt{3}$, and colloidal particle $2$ at
$x,y,z=R/2\sqrt{3}$. In order to keep the electro-neutrality, $2Z_{bare}$
microions of
charge $q$ are also present in the simulation box. 
If salt is added to the system, then $L^3
\rho_{S}$ microions of charge $q$ and $L^3 \rho_{S}$ microions of charge $-q$
are included inside the box. 
The total number of microions in the system is then $N=2L^3
\rho_{S}+2Z_{bare}$.
The radii of all the ions are set to $2$\AA.
The usual Coulomb potential is considered between all the charged species.
The total energy used in the MC simulations is:
\begin{equation}
\frac{\beta}{\lambda_B} E=\sum_{i=1}^{N-1}\sum_{j=i+1}^{N}
\frac{z_i^2}{r_{ij}}-\sum_{i=1}^{N} \frac{Z_{bare} z_i }{r_{1i}}-\sum_{i=1}^{N}
\frac{Z_{bare} z_i }{r_{2i}}
\end{equation}
where $z_i$ is the charge valence of the ion $i$~($+1$ or $-1$), $r_{ij}$ is the
distance between two ions $i$ and $j$, $r_{1i}$ and $r_{2i}$ are the distances 
between the ion $i$ and the colloidal particles $1$ and $2$, respectively. 
Since we consider
periodic boundary conditions, the Ewald summation technique is
employed~\cite{allen}. The equilibration is achieved after $2.5 \times 10^3$
simulation steps per particle; every $100$ movements per particle an
uncorrelated state is saved. The mean force is calculated using $1 \times 10^4$
uncorrelated configurations.

The average electrostatic force on a colloidal particle (positive force
corresponds to repulsion), along the diagonal
direction is
\begin{equation}
\bar F_e(R) = \left<\sum_{i=1}^{N} \frac{Z_{bare} z_i}{2} \left(
\frac{\cos{\theta_{1i}}}{r_{i1}^2} + \frac{\cos{\theta_{2i}}}{r_{i2}^2} \right)
\right> + \frac{Z_{bare}^2}{R^2} \ ,
\end{equation}
where $\bar F_e(R) = \frac{\beta}{\lambda_B} F_e(R)$, $\theta_{1i}$ and
$\theta_{2i}$ are the angles between the diagonal and the line connecting the
particle $i$ to the colloid $1$ and the colloid $2$, respectively.  These
distances are measured from
the diagonal in the counterclockwise, for particle $1$, and in the clockwise,
for 
particle $2$ direction,
respectively. The Ewald technique is used to calculate the electrostatic forces.
Besides the average electrostatic
force, there is also an entropic depletion force which must be taken into
account.
To do this we use the method of 
Wu~et~al.~\cite{WuBr99}, which consists of a small displacement of the colloidal
particles along
the diagonal (while the microions remain in a fixed configuration) 
in order to count the resulting overlaps between
the colloidal particle and the microions. This entropic force can be expressed
as
\begin{equation}
\bar F_d(R) = \frac{\left< N_1^c\right>-\left< N_1^f\right> + \left<
N_2^c\right>-\left< N_2^f\right>}{2 \Delta R \lambda_B} \ ,
\end{equation}
where $\bar F_d(R) = \frac{\beta}{\lambda_B} F_d(R)$, $N_1^c$ is
the number of overlaps of colloidal particle $1$ with the microions (both
anions and
cation), after a small
displacement $\Delta R$ ($\approx 1$\AA) in the direction of the 
colloidal particle $2$ (superscript c stands for closer)  and $N_1^f$, is the number of overlaps
after a displacement $\Delta R$ in the opposite direction (superscript f stands for farther). 
Similarly $N_2^c$ and
$N_2^f$, are the number of overlaps of colloidal particle $2$ with the
microions
after a displacement $\Delta R$ in the direction of the colloidal particle $1$ and in the
opposite direction respectively. The effective pair
potentials can then be calculated by integrating the 
mean force, $-\lambda_B\int_{R_{max}}^{R} dR' \left[ \bar
F_e(R') + \bar F_d(R') \right]$, where $R_{max}$ is the reference 
distance at which the interaction
between the two colloidal particles is negligible.

                         \section{Results}

We are now in a position to compare the thermodynamic predictions from
Eqs.~(\ref{eq6})
and ~(\ref{KB1}). To this end, the OZ equation is numerically solved using the
\textit{hipernetted-chain} (HNC) closure:
\begin{equation}
\label{eq10}
c(\mathbf{r})=h(\mathbf{r})-\log(h(\mathbf{r})+1)-\beta u(\mathbf{r}).
\end{equation} 
This closure is  known to be very accurate for Yukawa-like pair potentials
\cite{Hansen,Rev_Att}. For a
given reservoir salt concentration $\rho_{s}$ and volume fraction $\eta$, the
pair potential is given by (\ref{eq1}), with the effective charge 
calculated using the RJM.

\begin{figure}

\centering

\subfigure
{\includegraphics[width=4.2cm,height=4cm]{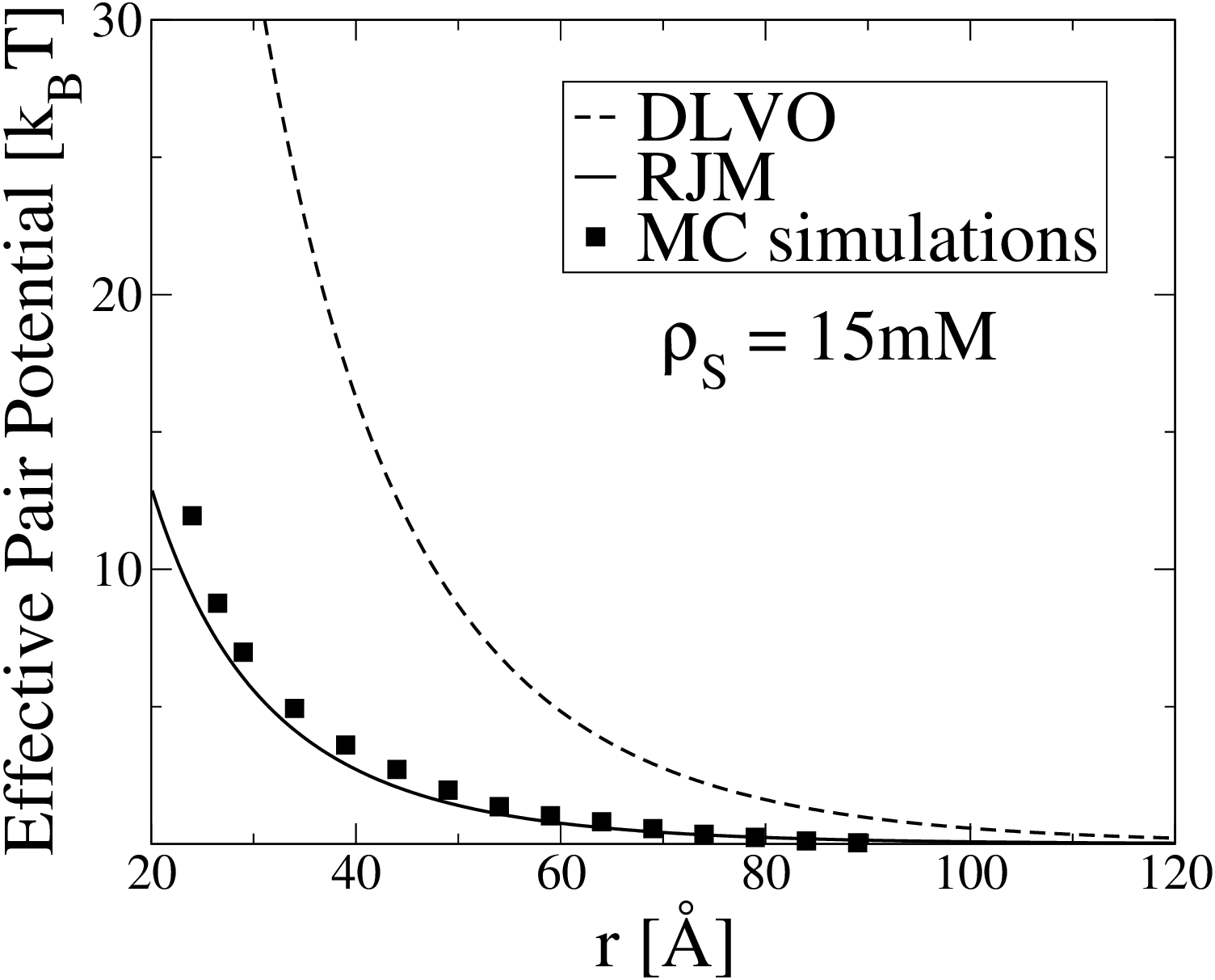}}
\vspace{0.3cm}
\subfigure{\includegraphics[width=4.2cm,height=4cm]{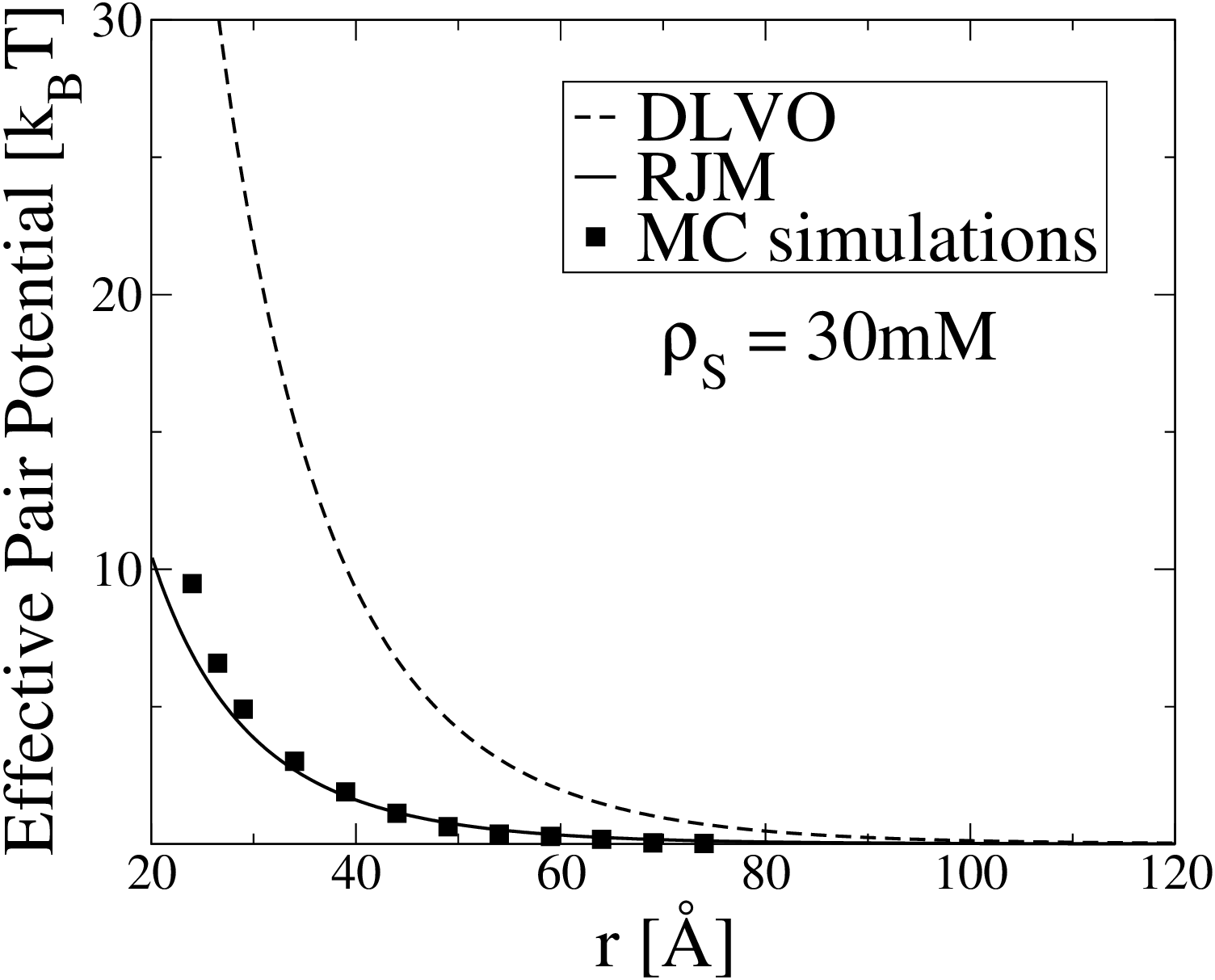}
}

\caption{Effective pair potentials calculated using the MC simulations (squares),
bare DLVO pair potential (dashed line) and DLVO with RJM effective parameters
(solid line) for $Z_{bare}=20$, $a=10$\AA\ and $\lambda_{B}=7.2$\AA.}
\label{fig1}
\end{figure}

In order to test the accuracy of the effective pair potential predicted by the
RJM, in Fig.~\ref{fig1}
we compare it with the results of the Monte
Carlo simulations.
As can be seen, the DLVO pair potential with the bare colloidal charge 
considerably overestimate the effective colloid-colloid interaction.
On the other hand, the pair-potential predicted by the RJM agrees well with the
MC 
simulations. Near the 
colloidal surface, however, a small deviation from the Yukawa functional form 
is evident.
These non-linear screening effects are a consequence of electrostatic
correlations
between the counterions near the colloidal surface. 

\begin{figure}
\centering
\subfigure
{\includegraphics[width=4cm,height=4cm]{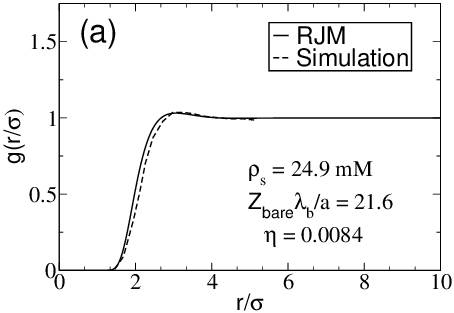}}
\vspace{0.3cm}
\subfigure{\includegraphics[width=4cm,height=4cm]{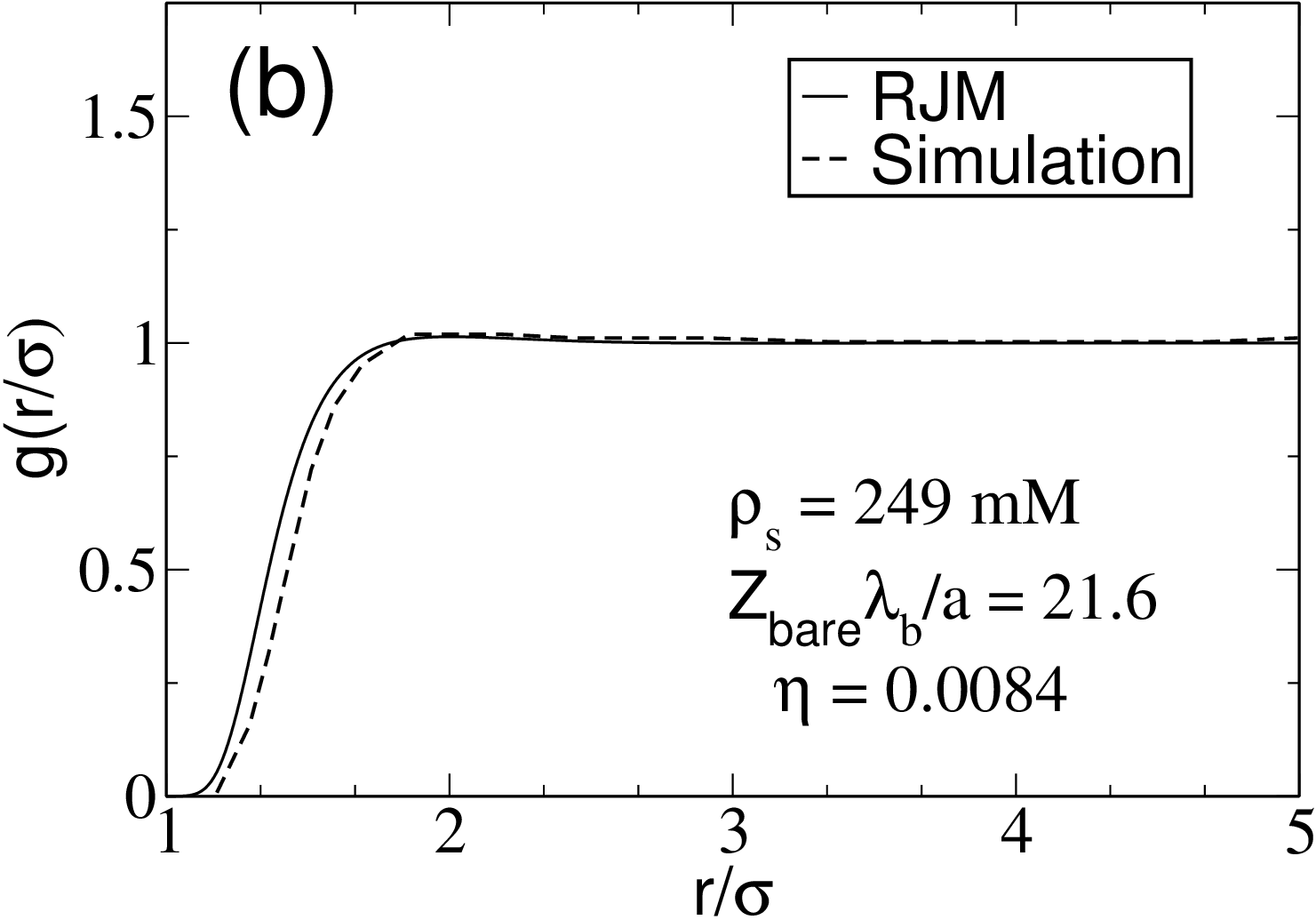}
}
\caption{Colloid-colloid pair correlation functions obtained using the MC
simulations (Ref.47) and the RJM-OZ approach, for a)
$\rho_{s}=24.9$ mM and b) $\rho_{s}=249$ mM. In both cases, the bare charge is
$Z_{bare}\lambda_{B}/a=21.6$, and the volume fraction is
$\eta=0.0084$.}
\label{fig2}
\end{figure}

In Fig.~\ref{fig2}, the colloid-colloid pair correlation function $g(r)$
calculated using 
the RJM and the HNC integral equation, is compared with the results of 
the MC simulations
\cite{Lob_03}  
in the high salt concentration regime. 
Again, we see a good agreement between the theory and the
simulations. 

\begin{figure}
\centering
\subfigure
{\includegraphics[width=4cm,height=4cm]{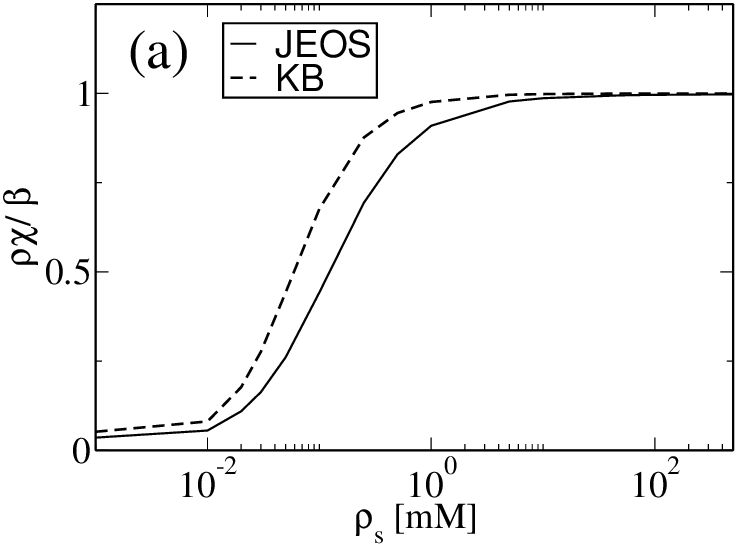}}
\subfigure{\includegraphics[width=4cm,height=4cm]{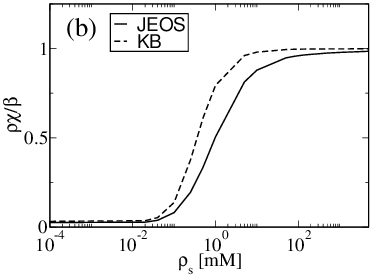}}
\subfigure{\includegraphics[width=4cm,height=4cm]{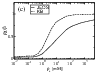}}
\vspace{0.3cm}
\subfigure{\includegraphics[width=4cm,height=4cm]{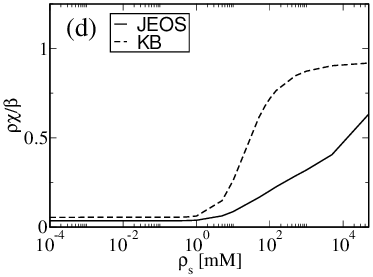}}
\caption{Reduced osmotic compressibility $\tilde{\chi}\equiv \rho\chi/\beta$ as
a function of the reservoir salt concentration $\rho_{s}$ for a colloidal particles of
radius $a=30 $\AA  and bare charge $Z=1000$. The colloidal volume fractions are: 
a) $\eta=10^{-5}$, b) $\eta=10^{-4}$, c)  $\eta=10^{-3}$ and d)
$\eta=10^{-2}$. We see a dramatic discrepancy between the predictions of the
JEOS (solid lines) and the Kirkwood-Buff fluctuation theory (dashed lines),
especially at
intermediate salt concentrations and high volume fractions.}
\label{fig3}
\end{figure}

Figs.~\ref{fig1} and \ref{fig2} show that
the effective
charges calculated using the RJM are able to correctly predict both the pair
interactions and the structural properties of colloidal suspensions containing
added electrolyte. We next check if this good agreement also
extends to the thermodynamic functions.  Unfortunately, very
quickly we run into difficulties.  We find that for the intermediate
salt concentrations, the osmotic compressibility calculated using the KB fluctuation 
relation Eq.~(\ref{KB1}) strongly
deviates from the
one calculated using the RJM equation of state (JEOS), Eq.~(\ref{JEOS}). The
discrepancy between the two routes can be clearly seen in Fig.~\ref{fig3}, which
shows
the osmotic compressibility $\chi_{osm}$ as a function of the reservoir salt
concentration $\rho_{s}$, for colloidal particles of bare charge 
$Z=1000$ and various  
volume fractions. Although both routes agrees in the low-salt and high-salt
regimes,
there are strong deviations at intermediate salt concentrations. Furthermore,
as 
the colloidal
concentration increases, the discrepancy between the two thermodynamic routes
becomes
stronger.  At low volume
fractions and high salt concentration, both routes approach the correct ideal
gas limit $\chi_{osm}\approx1$, when strong screening makes the 
system to behave
as a dilute suspension of hard spheres.

\begin{figure}
\centering
\subfigure
{\includegraphics[width=4cm,height=4cm]{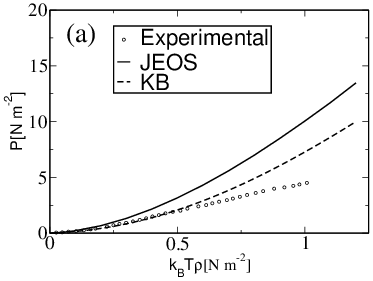}}
\subfigure{\includegraphics[width=4cm,height=4cm]{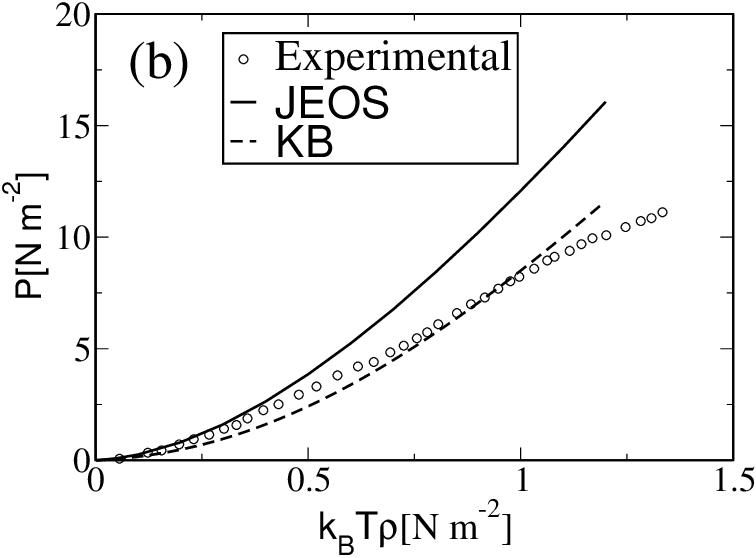}}
\caption{Comparison between the osmotic pressure calculated using the JEOS,
(solid line) and using the explicit integration of Eq.~(\ref{KB1}) (dashed line)
with the experimental results reported in Ref. 48. The reservoir salt
concentration is $\rho_{s}=8\mu$M, while the Bjerrum length is
$\lambda_{B}=2.38$ nm, the colloidal radius is $a=21.9$ nm, and colloidal
charges are: a) $Z=34$ and b) $Z=40$.}
\label{fig4}
\end{figure}

The question that arises then is: Which thermodynamic route is more reliable?
Unfortunately the answer is not very clear.  Due to the difficulty of 
performing large scale simulations on suspensions
containing electrolyte, there is very little data available to us to
answer this question.  Furthermore,  
there is also a scarcity of the experimental data dealing with
osmotic properties of charged colloidal suspensions.  
In Fig.~\ref{fig4}, we compare both the osmotic pressure calculated using the
JEOS and the KB 
fluctuation theory with the experimental measurements of Rasa et al. \cite{Ras_05}.
Neither one of the thermodynamic routes seems to be able to accurately describe
this
experimental data. Most likely  this is a
consequence  of the 
strong electrostatic correlations between the ions resulting from the use of a
low dielectric solvent 
by Rasa et al. Nevertheless, the fluctuation
route seems to give results in a closer agreement with the 
experimental data than the JEOS.  
This suggest that for the RJM the fluctuation route
might be more reliable for calculating the thermodynamic functions. 
We will now explore the 
possible causes of  
the discrepancy between the two
thermodynamic routes.

\subsection{Colloid-colloid correlations}

One possibility is that the discrepancy observed in Fig.~\ref{fig3} is due to
the way that colloidal correlations
enter into the theory in the two thermodynamic routes \cite{Bell_05,Dob_06}.
Indeed, while
the colloid-colloid correlations are neglected in the JEOS, they  contribute
to the osmotic pressure calculated using the KB formalism Eq.~(\ref{KB1}),
since the HNC equation \cite{Bell_05} used to obtain the correlation function
takes into account colloidal hard-cores. The colloid-colloid 
repulsion is particularly important for large volume fractions and  high-salt 
concentration ($Z\rho/2\rho_{s}\ll$1), when ionic contribution to
osmotic pressure is small. To asses the relevance of these
correlations, we can add to the JEOS the excess colloidal virial pressure,
\begin{equation}
\label{eq11}
\beta P^{ex} =
-\dfrac{2\pi\rho^{2}}{3}\int_{0}^{\infty}g(r)r^{3}\dfrac{d\beta u}{dr}dr
+2\pi\rho^{2}V\int_{0}^{\infty}g(r)\dfrac{\partial \beta u}{\partial V}r^{2}dr, 
\end{equation}
where $u(r)$ is the effective pair potential in the OCM description. The
first term
on the right hand side of this equation represents the standard excess virial
pressure for the one-component system, while the second term accounts
for the density dependent effective pair potential. This term 
is essential to reproduce the correct Debye-H\"uckel
limiting law in the infinite dilution limit \cite{Chan_01}. 
Substitution of Eq.~(\ref{eq1}) into Eq.~(\ref{eq11}) produces the following
expression for the excess pressure,
\begin{equation}
\label{eq12}
\begin{split}
\beta P^{ex} = \dfrac{2\pi\rho^2}{3}\int_{0}^{\infty}{g(r)\beta u(r)(\kappa r
+1)r^{2}dr} 
\\
-\pi\rho^{2}\int_{0}^{\infty}g(r)\beta u(r)\left(\kappa
r-\dfrac{2(\kappa a)^{2}}{1+\kappa a}\right)r^{2} dr,
\end{split}
\end{equation}
where $u(r)$ is the effective colloidal pair potential, Eq.~\ref{eq1}.
For all the parameters studied here, however, we find that
$\arrowvert P^{ex}\arrowvert\ll P_{Jell}$, 
and the effect of colloidal correlations
is too small to account for the strong discrepancy observed in
Fig.~\ref{fig3}.  

\subsection{Ion-ion correlations}

As the salt concentration increases, the mean distance between the 
cations and anions
becomes smaller, leading to strong
inter-ionic correlations.  Such correlations are completely
ignored by the mean-field Poisson-Boltzmann equation, which is  the
basis of the RJM.  Indeed, in the absence coions (and for monovalent
counterions), 
the RJM model was found to provide an excellent account of both thermodynamic
and structural properties of charged suspensions \cite{Emm_04,Sal_07,Cas_06,Col_09}. 
This good accuracy of the model
is the result of large characteristic distance between the counterions 
inside a salt-free suspension.  
On the other hand, presence of salt leads to strong 
cation-anion correlations neglected 
in the RJM.   

In order to explore the influence of inter-ionic correlations on the osmotic
pressure in a colloidal suspension, we modify the JEOS by
adding the correlational Debye-H\"uckel contribution \cite{Rev_Lev,Lev_96}, 
\begin{equation}
\label{eq13}
\beta P^{ex}=-\dfrac{\kappa^{3}}{24\pi}.
\end{equation}

\begin{figure}
\centering
\subfigure
{\includegraphics[width=4cm,height=4cm]{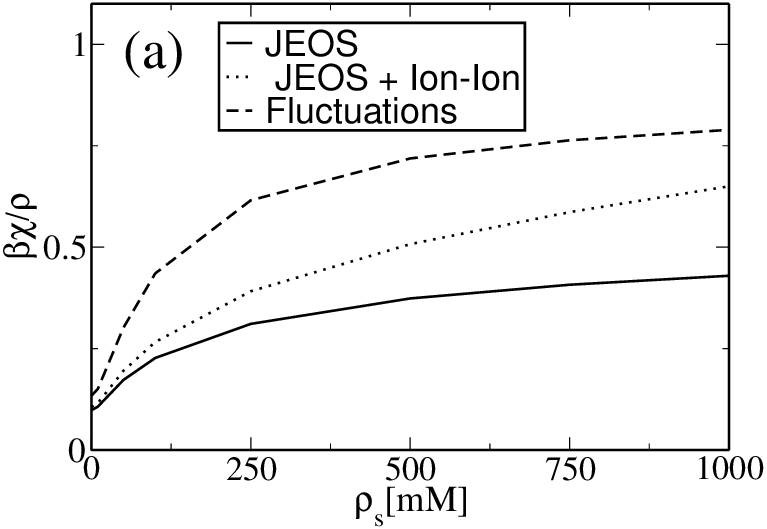}}
\subfigure{\includegraphics[width=4cm,height=4cm]{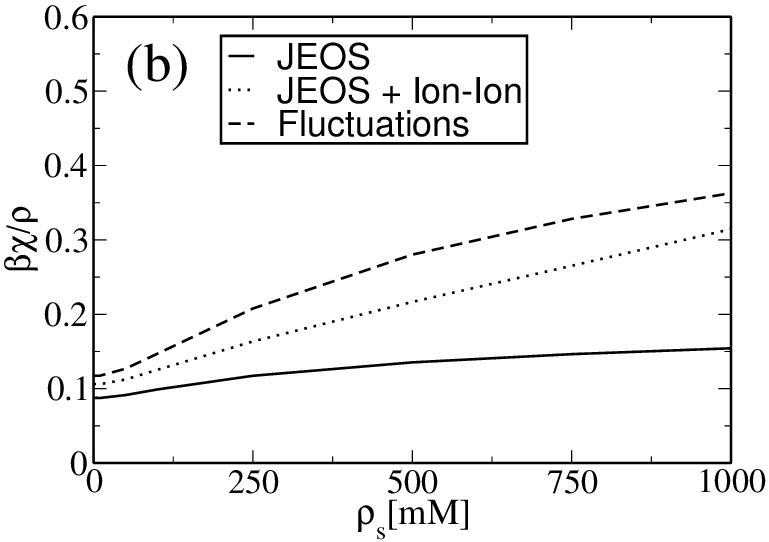}}
\caption{Comparison between the osmotic compressibilities calculated using the
JEOS Eq.~\ref{JEOS} (solid lines), using the JEOS with explicit ionic
correlations Eqs.~(\ref{JEOS}) and (\ref{eq13}), and using the the KB fluctuation
theory, Eq.~(\ref{KB1}). The radius of colloidal particles is $a=10$\AA, 
the bare colloidal
charge is $Z=1000$, and the Bjerrum length is $\lambda_{b}=7.2$\AA. The volume
fractions are: a) $\eta=0.01$ and b) $\eta=0.05$.   }
\label{fig DH}
\end{figure}

Figure \ref{fig DH} shows the osmotic compressibilities resulting from 
addition of Eq.~\ref{eq13} to the JEOS, Eq.~\ref{JEOS}. 
As can be seen from this figure,
incorporation of ionic correlations even at this leading-order level, 
significantly improves the agreement between the two thermodynamic routes,
especially at large  colloidal volume fractions. 
This simple calculation 
suggests that the
thermodynamic consistency of the RJM can be restored by incorporating the
inter-ionic correlations into the RJM.  Unfortunately, at the moment, it is not
clear how the inter-ionic correlations can be included into the RJM in a fully
self-consistent fashion.  This will be the subject of future research.

%
%
                         \section{Summary and Conclusions}                      
                                                                           
%
%

We have reported an inconsistency arising when different routes are
employed to calculate the thermodynamic functions in the 
the RJM.  The discrepancies are particularly strong
at moderate salt concentrations. On the other hand, comparing
the predictions of the  RJM with the MC simulations, 
we see that the model accurately accounts for the effective pair
interactions and
the colloidal correlation functions, even for suspensions containing
electrolyte. 

Thermodynamic inconsistency between different routes is not particular to
the 
RJM
and is found for many other systems \cite{Jay_67}.  Even for a
Debye-H\"uckel electrolyte, the osmotic compressibility calculated via the PM
virial
equation is quite different from the predictions of the fluctuation theory
\cite{Rev_Bell}. In
these cases, MC simulations are particularly helpful to choose the more
accurate route to thermodynamics \cite{Jay_67}.
Unfortunately, simulations of charged colloidal suspensions at even moderate
salt
concentrations are still too computationally demanding while the experimental
data is still very scarce.  The experimental and the simulational data available
to us
seems to  indicate 
that KB fluctuation relations provide a more reliable route to 
thermodynamics
of the RJM.  The KB route seems to partially account for the inter-ionic
correlations
which are completely neglected by the JEOS. These correlations are negligible
in the absence of coions, they however become relevant when 
salt concentration increases and the characteristic distance between the cations
and the anions becomes small \cite{Rev_Lev}. In Section IV (B), 
we showed that even a simple 
incorporation of the DH contribution to the osmotic pressure already  
brings the JEOS and the fluctuation results into a
closer agreement. A fully self-consistent incorporation of ionic correlations
into
the RJM requires, however, development of a new methodology closer in spirit to
the
density functional theory. The work in this direction
is now in progress.

\section{Acknowledgments}

This work was partially supported by the CNPq, FAPERGS, INCT-FCx, and by the 
US-AFOSR under the grant FA9550-09-1-0283.

\bibliographystyle{prsty}
\bibliography{paper}

\begin{thebibliography}{10}

\bibitem{Rev_Lev}
Y. Levin, Rep. Prog. Phys. {\bf 65},  1577  (2002).

\bibitem{Rev_Mes}
R. Messina, J. Phys.: Condens. Matter {\bf 21},  113102  (2009).

\bibitem{Rev_Bell}
L. Belloni, J. Phys.: Condens. Matter {\bf 12},  R549  (2000).

\bibitem{Mah_00}
K. Mahdi and M.~O. de~la Cruz, Macromolecules {\bf 33},  7649  (2000).

\bibitem{Die_01}
A. Diehl, M.~C. Barbosa, and Y. Levin, Europhys. Lett. {\bf 53},  86  (2001).

\bibitem{Lev_03}
Y. Levin, E. Trizac, and L. Bocquet, J. Phys.: Cond. Mat. {\bf 15},  S3523
  (2003).

\bibitem{Den_07}
A.~R. Denton, Phys. Rev. E {\bf 2007},  051401  (76).

\bibitem{Ben_07}
B. Lu and A.~R. Denton, Phys. Rev. E {\bf 75},  061403  (2007).

\bibitem{Rev_Han}
J.-P. Hansen and H. Löwen, Annu. Rev. Phys. Chem. {\bf 51},  209  (2000).

\bibitem{Ver_11}
G. Vernizzi, G.~I. Guerrero-Garcia, and M.~O. de~la Cruz, Phys. Rev. E {\bf
  84},  016707  (2011).

\bibitem{Tam_98}
M. Tamashiro, Y. Levin, and M. Barbosa, Eur. Phys. J. B {\bf 1},  337  (1998).

\bibitem{Des_02}
M. Deserno and H.~H. von Grünberg, Phys. Rev. E {\bf 66},  011401  (2002).

\bibitem{Bell_98}
L. Belloni, Colloids and Surfaces A {\bf 140},  227  (1998).

\bibitem{Gon_08}
P. Gonzalez-Mozuleos and M.~O. de~la Cruz, Physica A {\bf 387},  5362  (2008).

\bibitem{Rev_Lik}
C.~N. Likos, Physics Reports {\bf 348},  247  (2001).

\bibitem{Rev_Den}
{A. Denton}, {\em in Nanostructured Soft Matter: Experiment, Theory, Simulation
  and Perspectives} (Springer, Dordrecht, 2007).

\bibitem{Den_03}
A.~R. Denton, Phys. Rev. E {\bf 67},  011804  (2003).

\bibitem{Sti_02}
F.~H. Stillinger, H. Sakai, and S. Torquato, J. Chem. Phys. {\bf 117},  288
  (2002).

\bibitem{Louis_02}
A.~A. Louis, J. Phys.: Condens. Matter {\bf 14},  9187  (2002).

\bibitem{Dijk_00}
M. Dijkstra, R. van Roij, and R. Evans, J. Chem. Phys. {\bf 113},  4799
  (2000).

\bibitem{Emm_07}
E. Trizac {\it et~al.}, Phys. Rev. E {\bf 75},  011401  (2007).

\bibitem{DLVO1}
B.~V. Derjaguin and L. Landau, Acta Physicochim. URSS {\bf 14},  633  (1941).

\bibitem{DLVO2}
{E. J. Verwey and J. T. G. Overbeek}, {\em Theory of the Stability of Lyophobic
  Colloids} (Elsevier, Amsterdam, 1948).

\bibitem{Bas&Roij_06}
B. Zoetekouw and P.~R.~E. R.~van Roij, Phys. Rev. E {\bf 73},  021403  (2006).

\bibitem{Chan_01}
D.~Y.~C. Chan, P. Linse, and S.~N. Petris, Langmuir {\bf 17},  4202  (2001).

\bibitem{SL}
F. Stillinger and R. Lovett, J. Chem. Phys. {\bf 49},  1991  (1968).

\bibitem{Fri_70}
H.~L. Friedman and P.~S. Ramanathan, J. Phys. Chem. {\bf 74},  3756  (1970).

\bibitem{Kus_87}
P.~G. Kusalik and G.~N. Patey, J. Chem. Phys. {\bf 86},  5110  (1987).

\bibitem{Emm_04}
E. Trizac and Y. Levin, Phys. Rev. E {\bf 2004},  031403  (69).

\bibitem{Sal_07}
S. Pianegonda, E. Trizac, and Y. Levin, J. Chem. Phys. {\bf 126},  014702
  (2007).

\bibitem{Hansen}
{ J.-P Hansen and I.R. McDonald}, {\em Theory os Simple Liquids} (Academic
  Press, New York, 1976).

\bibitem{KB}
J.~G. Kirkwood and F.~P. Buff, J. Chem. Phys. {\bf 19},  774  (1951).

\bibitem{Cas_06}
R. Castaneda-Priego, L.~F. Rojas-Ochoa, V. Lobaskin, and J.~C. Mixteco-Sanchez,
  Phys. Rev. E {\bf 74},  051408  (2006).

\bibitem{Col_09}
T.~E. Colla, Y. Levin, and E. Trizac, J. Chem. Phys. {\bf 131},  074115
  (2009).

\bibitem{Col_10}
T.~E. Colla and Y. Levin, J. Chem. Phys. {\bf 133},  234105  (2010).

\bibitem{Fal_11}
J.~M. Falcon-Gonzalez and R. Castañeda-Priego, Phys. Rev. E {\bf 83},  041401
  (2011).

\bibitem{Fal_10}
J.~M. Falcon-Gonzalez and R.~C. Priego, J. Chem. Phys. {\bf 133},  216101
  (2010).

\bibitem{Wen_82}
H. Wennerstrom, B. Jonsson, and P. Linse, J. Chem. Phys. {\bf 76},  4665
  (1982).

\bibitem{Des_01}
M. Deserno and C. Holm, in ``Electrostatic Effects in Soft Matter and
  Biophysics'', eds. C. Holm, P. Kekicheff, and R. Podgornik, NATO Science
  Series II - Mathematics, Physics and Chemistry {\bf 46},    (2001).

\bibitem{Car_81}
S.~L. Carnie and D.~Y.~C. Chan, Chem. Phys. Lett. {\bf 77},  437  (1981).

\bibitem{Mar_55}
R.~A. Marcus, J. Chem. Phys. {\bf 23},  1057  (1955).

\bibitem{Donnan}
F.~G. Donnan, Chem. Rev. {\bf 1},  73  (1924).

\bibitem{Beh_97}
R. Behera, J. Chem. Phys. {\bf 108},  3373  (1997).

\bibitem{allen}
{M. P. Allen and D. J. Tildesley}, {\em Computer Simulations of Liquids}
  (Oxford University Press, Oxford, 1987).

\bibitem{WuBr99}
J.~Z. Wu, D. Bratko, H.~W. Blanch, and J.~M. Prausnitz, J. Chem. Phys. {\bf
  111},  7084  (1999).

\bibitem{Rev_Att}
P. Attard, Avd. Chem. Phys. {\bf 92},  1  (1996).

\bibitem{Lob_03}
V. Lobaskin and K. Qamhieh, J. Phys. Chem. B {\bf 107},  8022  (2003).

\bibitem{Ras_05}
M. Rasa {\it et~al.}, J. Phys.: Condens. Matter {\bf 17},  2293  (2005).

\bibitem{Bell_05}
L. Belloni, J. Chem. Phys. {\bf 123},  204705  (2005).

\bibitem{Dob_06}
J. Dobnikar, R. Castañeda-Priego, H. von Grünberg, and E. Trizac, New Journal
  of Physics {\bf 8},  277  (2006).

\bibitem{Lev_96}
Y. Levin and M. Fisher, Physica A {\bf 225},  164  (1996).

\bibitem{Jay_67}
J.~C. Rasaiah and H.~L. Friedman, J. Chem. Phys. {\bf 48},  2742  (1968).

\end{thebibliography}

\end{document}